# Control of Mechanical and Fracture Properties in Two-phase Materials Reinforced by Continuous, Irregular Networks

Tommaso Magrini*, Chelsea Fox, Adeline Wihardja, Athena Kolli, Chiara Daraio*

*Engineering and Applied Science, California Institute of Technology, Pasadena, CA 91125, USA*
*magrini@caltech.edu, *daraio@caltech.edu

**Abstract**

Composites with high strength and high fracture resistance are desirable for structural and protective applications. Most composites, however, suffer from poor damage tolerance and are prone to unpredictable fractures. Understanding the behavior of materials with an irregular reinforcement phase offers fundamental guidelines for tailoring their performance. Here, we study the fracture nucleation and propagation in two phase composites, as a function of the topology of their irregular microstructures. We use a stochastic algorithm to design the polymeric reinforcing network, achieving independent control of topology and geometry of the microstructure. By tuning the local connectivity of isodense tiles and their assembly into larger structures, we tailor the mechanical and fracture properties of the architected composites, at the local and global scale. Finally, combining different reinforcing networks into a spatially determined meso-scale assembly, we demonstrate how the spatial propagation of fractures in architected composite materials can be designed and controlled *a priori*.

**Introduction**

Composite materials offer many advantages over traditional materials, such as being lightweight while maintaining a high strength and stiffness,[1,2] but they suffer from lack of toughness and poor damage tolerance.[3–6] One way to improve their crack response is to tailor the reinforcing phase architecture.[7–10] Fiber reinforcements, for example, exploit crack bridging between fibers for toughening. Introducing fibers and other high-aspect-ratio reinforcing elements in the design of composite materials, often leads to direction-dependent mechanical properties and anisotropic fracture resistance.[11] Depending on the reinforcing elements' alignment direction, composites can be either toughened by high fracture energy dissipative mechanisms, such as fiber bridging and fiber pullout, or be subject to delamination fractures, which occur at the fiber-matrix interface.[11–14] On the contrary, randomly distributed inclusions, which primarily toughen the material through microcracking and secondary crack formation, often lead to composite materials with isotropic fracture properties.[15–17] Developing materials that use multiple toughening mechanisms, like bridging, deflecting, or even arresting the propagation of cracks, has potential to improve the amount of absorbed fracture energy. This was recently demonstrated in bioinspired architected composites, where the internal microstructure is finely tailored to control crack propagation behavior.[18,19] The combination of multiple toughening mechanisms can also be achieved by fabricating composite materials with irregular reinforcing networks[20,21]. Irregular microstructures are common in biological structural materials[22–25] and understanding their behavior during loading and fracture is relevant for the design of architected materials with tailored load-bearing performance. Irregular networks can control the fracture and toughening behavior of materials through the creation of meso-scale structures with different dimensions and orientations that cause multiple fracture nucleation and propagation events. Finally, reinforcing composites with irregular networks allows the creation of materials with direction-independent mechanical properties, a desirable feature in structural and load-bearing applications. Here, we describe how network coordination influences the global mechanical properties of two-phase materials, like strength, stiffness and energy dissipated during fracture, as well as the role of local mechanisms on fracture nucleation and propagation. Introducing desired irregular networks as composite reinforcement and achieving a fine control over their assembly across multiple lengthscales, from the micro- to the cm-scale, requires advances in both numerical design and manufacturing. In recent work, machine-learning and data-driven approaches were used to computationally design hierarchical architected materials.[26] Here, we employ algorithms that "grow" regular and irregular networks [27] for composite design and use multi-material additive manufacturing processes for fabrication.

**Design of irregular reinforcement**

To design the stiff reinforcement phase of our two-phase composites, we utilized the virtual growth algorithm (Supplementary Discussion 1), which tessellates a set of bimaterial tiles on a discretized spatial grid, following a set of adjacency rules[22]. We used a combination of 2-coordinated tiles ([L] and [-]) and 3-coordinated tiles ([T]) and ensured that each tile had the same volume fraction of stiff reinforcing phase and soft matrix phase (Figure 1a, left). We combined these tiles to generate composites with a stiff reinforcing random network (white) and a soft elastomeric matrix (black) (Figure 1a, right). The virtual growth algorithm ensures continuity between the two phases through modifiable connectivity rules (Figure SI 1). Depending on the relative composition of 2- and 3-coordinated tiles, the virtual growth algorithm creates various composites with the same volume fraction of reinforcement, but a large ternary design space (Figure 1b). We expect the shape and directional tile connectivity to influence the local deformation mechanisms accessible within the clusters, with [L] shaped tiles showing bending-dominated local deformations and straight [-] tiles showing stretching-dominated behaviors.

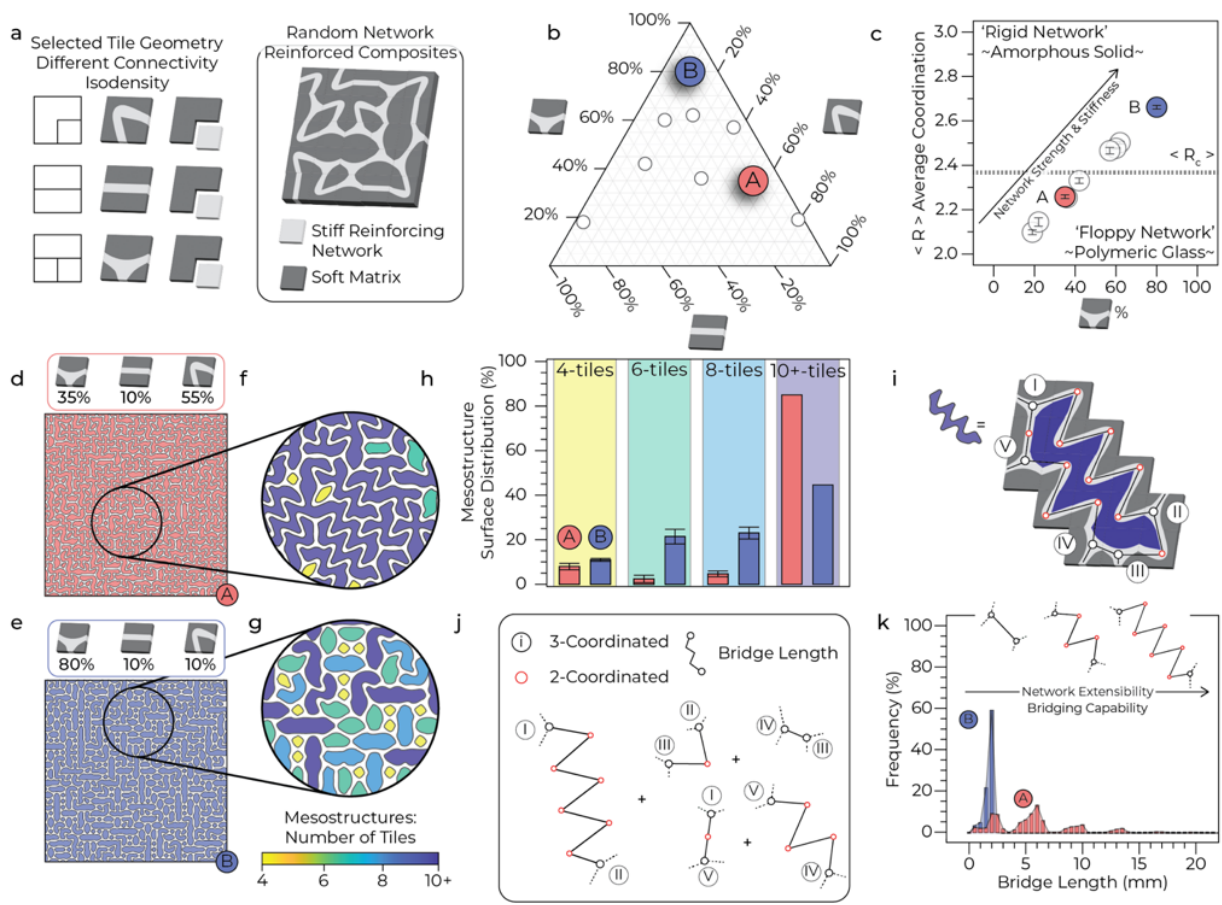

**Figure 1. Architecture of two-phase materials.** (a) Selected isodensity tile geometries and composite assembly. (b) Compositional design space in a ternary diagram. (A) and (B) architectures are represented by red and blue circles, respectively. (c) Average coordination <R> as a function of [T] tiles content. (A) and (B) reinforcing networks are represented by red and blue circles, respectively. (d, e) Representative (A) and (B) architectures (d and e, respectively). (f, g) Close-up view of meso-structures that populate (A) and (B) architectures in (d) and (e) respectively. Yellow, green, cyan, and blue represent 4, 6, 8, and 10+ tiles meso-structures, respectively. (h) Meso-structures distribution in (A) (red bars) and (B) architectures (blue bars). (i) Example of meso-structure with labeled coordination and bridges. (j) Expanded version of (i). (k) Comparison of bridge length and their frequency for (A) and (B) architectures (red and blue, respectively).

**Network characterization**

We evaluate the properties of the reinforcing networks using frameworks developed to describe covalent random networks (Supplementary Discussion 2), at two hierarchical scales. At the global scale, we evaluate the average coordination of the materials at constant density, and at the local scale, we analyze how growth rules affect the formation of characteristic meso-structures. We evaluate the average coordination <R> in the reinforcing networks, accounting for the presence of dangling bonds, unconnected ligaments at the network edges (Figure 1c).[28,29] Scaling linearly with the volume fraction of 3-coordinated tiles, we expect <R> to influence the global mechanical properties, like strength and stiffness, as reported in other amorphous materials systems.[30–32] To understand the effect of the reinforcing network architecture on the composite properties, we compare two different compositions with significantly different average coordination: (A)-networks (35 [T], 10 [-], 55 [L]), dominated by 2-coordinated tiles and floppy modes; and (B)-networks (80 [T], 10 [-], 10 [L]), dominated by 3-coordinated tiles and that are purely rigid (figure 1b and figure 1c, red and blue circles respectively).

Despite having the same reinforcing and matrix phase volume fractions, (A)- and (B)-network reinforced composites (NRCs) form different local meso-structures, defined as the matrix domains enclosed by reinforcing network (Figure 1d and 1e). While the average coordination of the reinforcing network explains the global mechanical behavior of the materials, studying the meso-structures that pattern each composite is key to understand their local properties. First, the meso-structures are categorized and mapped based on size and number of constitutive tiles (Figure 1f and 1g). Then, their surface distribution is used to indicate the texture of (A)- and (B)-NRCs (Figure 1h). Additionally, the number density of each meso-structure (Figure SI 2), their angle of orientation (Figure SI 3), and the effect that small meso-structures have on their surroundings (Figure SI 4) are important descriptors of these architected composites.

We characterize the reinforcing networks by drawing parallels with the concept of network bridges, often used in studying of the mechanical performance of covalent random networks.[28,29] A bridge (black solid lines, Figure 1i-1j) connects two 3-coordinated tiles, considered anchored in the network (I-V white circles, Figure 1i-1j). It was demonstrated that a bridge composed of 6 or more 2-coordinated tiles (red circles, Figure 1i-1j) forms a floppy region within the network.[28,29] The presence of floppy domains in a stiff, yet deformable, reinforcing network influences the local mechanical composite performance, resulting in a globally more extensible and deformable material (Figure SI 5). In this context, the presence of an incompressible matrix phase is important to prevent large bridge deformations. Because of the different content of 3-coordinated tiles, (A)-NRCs display a multi-modal distribution of bridge lengths which are significantly longer than those of (B)-NRCs (Figure 1k).

**Mechanical Properties**

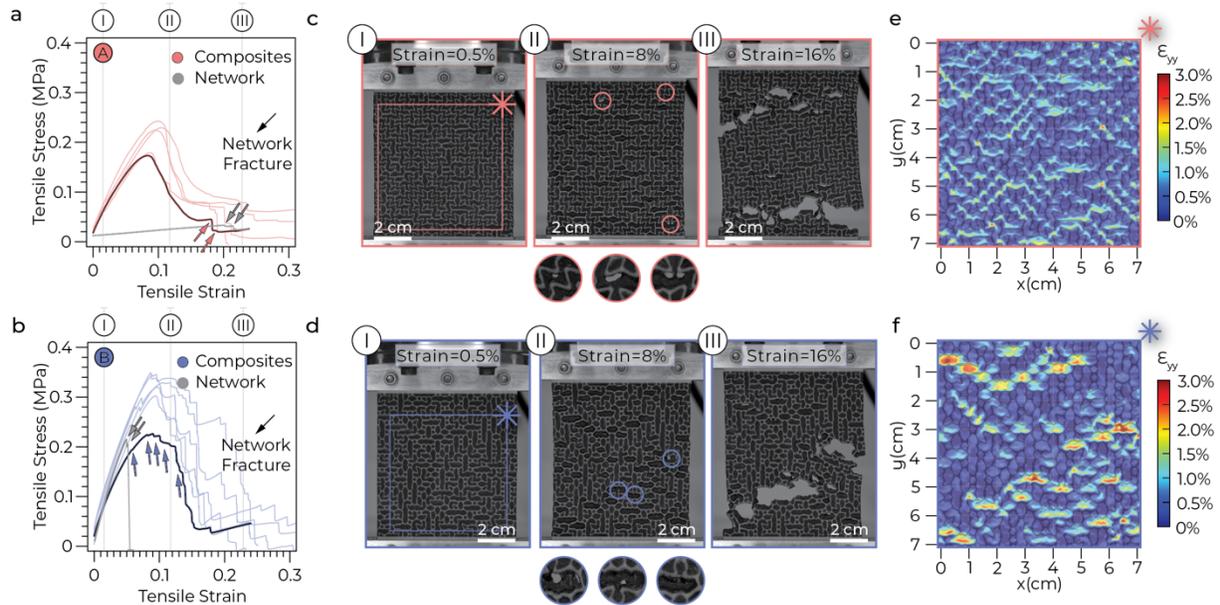

**Figure 2**. **Mechanical characterization of composites.** (a, b) Engineering stress-strain curves recorded during uniaxial tension tests on plate geometries of (A)-NRCs and (B)-NRCs (red solid lines in a, blue solid lines in b, respectively). The solid black lines in (a,b) represent the response of samples photographed in (c, d), respectively. The solid gray lines in (a,b) represent the response of the same (A)-NRC and (B)-NRC samples, without the matrix phase. Fracture events in the reinforcing phase of (A)-NRCs and (B)-NRCs are indicated by red and blue arrows in a and b, respectively, and in the reinforcing networks by grey arrows (see also Figure SI 6). (c, d) Fracture evolution in representative specimens of (A)-NRC and (B)-NRC, respectively. The circles indicate the locations within the specimens that display the signs of voids growth (circles in c and d, frame II and insets in c and d, frame II, bottom). (e, f) Digital image correlation (DIC) maps of the representative specimens of (A)-NRC and (B)-NRC recorded at 0.5% strain (e and f respectively). The DIC maps refer to the areas of specimens highlighted by (*) in frame I of (c) and (d).

Although (A)- and (B)-NRCs have the same volume fraction of reinforcement and matrix phases, the difference in average coordination and bridge length and different meso-structure populations influence the mechanical properties at both global and local scales. To measure experimentally the mechanical properties of the chosen architectures, we additively manufactured composite samples using a polyjet printer (Stratasys Objet500 Connex3). Recent studies have focused on determining experimentally the mechanical and physical properties of objects printed by polyjet printing and shed light on the relationship between the printing parameters and the final performance of the part.[33–35] In our study, a stiff viscoelastic resin (VeroWhite Polyjet Resin) and a soft elastomeric resin (TangoBlack Polyjet Resin) were chosen for the reinforcing phase and matrix phase, respectively. Both resins are commercially available, and their constitutive properties fall within ranges reported in literature (Figure SI 7).[18,36–38] We combined these two materials in a polymer composite with a volume fraction of reinforcing phase of 0.3. At this volume fraction, we observed that the composites display a desired tradeoff between rigidity and extensibility (Figure SI 8), while the reinforcing network thickness is one order of magnitude larger than the polyjet printer resolution limit (Figure SI 9). To characterize their mechanical response, we performed plate tension experiments and confirmed that at the global scale, the purely rigid-like (B)-networks achieve higher strength and higher stiffness than the (A)-networks (Figure 2a and 2b).

Despite a significant difference in the global mechanical properties, the composites display similarities in the local scale mechanisms that determine the initiation and propagation of fractures. Due to the remarkable adhesion properties between the two resins used in this study[39], fracture initiation does not occur at the interface between the matrix and the reinforcing network, in either pristine or pre-notched samples, but within the matrix (Figure SI 7). Void nucleation in the matrix phase initiates the composite fracture process, similar to the ductile fracture of metals.[40] Void formation is followed by matrix detachment from the reinforcing network, resulting in steady void growth (Figure 2c and 2d, I to III respectively). In this propagation phase, the void growth and coalescence are hindered by the reinforcing network bridges, which elongate as the sample undergoes tensile loading. Thus, the average bridge length and extensibility before rupture become paramount, as these characteristics predict the strain of the reinforcing network before failure (Supplementary Discussion 3). After the sequential failure of the bridges (Figure 2a and 2b, red and blue arrows respectively), we observe the complete loss of composite integrity.

The local composite architecture become key during failure, as strain localization in selected meso-structures leads to fracture nucleation and growth, as confirmed by 2D Digital Image Correlation (DIC) at small strains (Figure 2e and 2f). Therefore, to design composites capable of dissipating the most fracture energy, one must act on both the global and local scale, tailoring the network rigidity and generating local meso-structures, to avoid localized strain fields. To achieve this, we modify the connectivity rules of the growth algorithm.

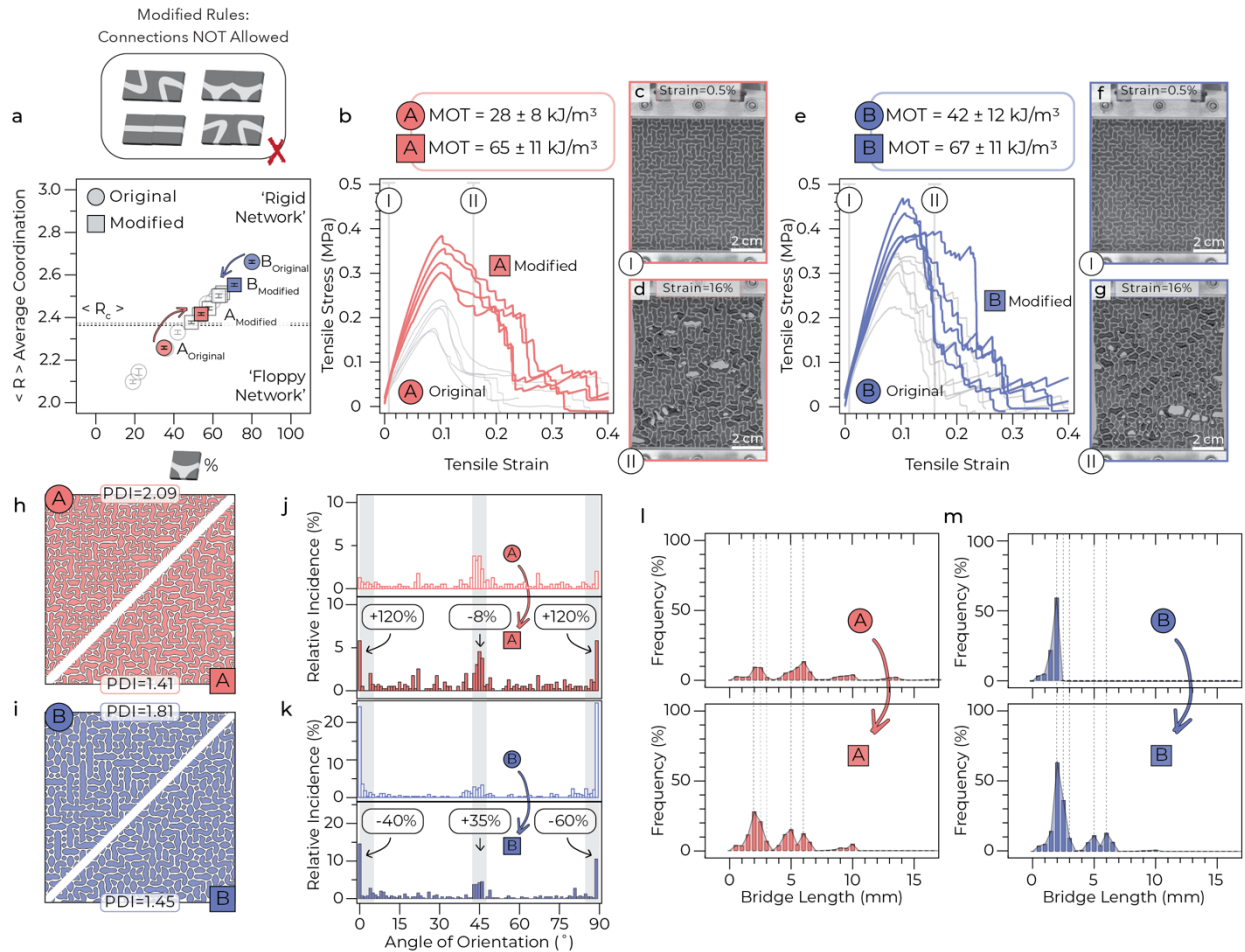

**Figure 3. Modified composites and their performance.** (a) Modifications of connectivity rules and average coordination number as a function of [T] tiles (top and bottom, respectively). (b) Engineering stress-strain diagram of Mod-(A)-NRCs (red solid lines) and of original (A)-NRCs (grey solid lines). The modulus of toughness (MOT) is reported for both composites at the top of the diagram. (c, d) Fracture evolution in representative Mod-(A)-NRCs at 0.5% and 16% strain (c and d, respectively). (e) Engineering stress-strain diagram of Mod-(B)-NRCs (blue solid lines) and of the original (B)-NRCs (grey solid lines). The MOT is reported for both composites at the top of the diagram. (f, g) Fracture evolution in representative Mod-(B)-NRC at 0.5% and 16% strain (f and g, respectively). (h, i) Modification of microstructure of (A)- and (B)-networks (h and i respectively) and measured polydispersity index (PDI) for each network. (j, k) Variation in relative meso-structure orientation distribution of (A) and (B)-networks (j and k respectively). (l) Frequency of bridge lengths for (A)- and Mod-(A)-networks (top and bottom, respectively). (m) Frequency of bridge lengths for (B)- and Mod-(B)-networks (top and bottom, respectively).

We changed the connectivity rules of a growth algorithm to increase energy dissipation during fracture in composites. By amending four tile adjacency rules (Figure 3a, top, Supplementary Discussion 4, Figure SI 10), we prevented the formation of large floppy domains, which increased network rigidity, stiffness, and strength. The modified networks displayed a purely rigid-like behavior, as shown by their higher average coordination than the original networks (Figure 3a, bottom). We tested the effect of the modified reinforcing networks on the composites' mechanical performance and fracture energy dissipation through plate tension experiments. As a result of their higher coordination, Mod-(A)-NRCs displayed higher ultimate tensile strength (UTS) and up to 60% increase in tensile stiffness (Figure 3b red and grey solid lines respectively), while Mod-(B)-NRCs had a 5% reduction in stiffness as a result of the slightly lower average coordination (Figure 3e blue and grey solid lines respectively). Although each composite begins failure at ~10% tensile strain, the modified designs' damage tolerance dramatically improved. At high tensile strain (up to ~16%), the Mod-NRCs carry a load of approximately 70-80% their UTS (Figure 3c-d and Figure 3f-g). As a comparison, their original counterparts at the same tensile strain had completely lost any load carrying capabilities, due to presence of sample-scale cracks and coalesced voids, resulting from the extensive failure of the reinforcing phase. Conventional calculations of the stress intensity factor and local stress concentration field require making assumptions based on continuum mechanics: for composite materials, the reinforcing feature sizes must be small compared to the size of the singularity zone, and the nonlinear damage must be confined to a small region within the singularity zone.[40] In our irregular composites these conditions are not satisfied: meso-structures sizes are in the order of several mm (Figure SI 2) and crack nucleation occurs in multiple locations within the microstructure (Figures 2c, 2d, 3d and 3g). In the present study, to highlight how these simple modifications to the reinforcing networks influence significantly the energy dissipated during fracture, we measured the modulus of toughness (MOT), taken as the area under the stress-strain curve. Modifying the reinforcing networks in (A) and (B) composites improved the total dissipated energy during fracture of up to ~130% and ~60% respectively (Figure 3b and 3e, top).

Considering global scale descriptors solely, like the average reinforcing network coordination, is insufficient to explain the higher strength of Mod-(B)-NRCs compared to (B)-NRCs. Thus, we evaluated the modified designs at the local scale, to investigate the effect that simple modifications of the connectivity rules had on the meso-structures. First, we notice by visual inspection that the modified composites (Figure 3h and 3i, bottom) have a significantly different internal structure than their original counterparts (Figure 3h and 3i, top). The modified architectures feature a more homogeneous distribution of meso-structures, which are quantified through the polydispersity index (PDI) (Figure 3h and 3i, Supplementary Discussion 5). The decrease in PDI by 33% for (A)-NRCs and by 20% for (B)-NRCs, confirms that more stringent connectivity rules homogenize and coarsen the meso-structures sizes (Figure SI 11). Furthermore, the modified composites feature meso-structures that display a more homogeneous angle of orientation with respect to their original counterparts (Figure 3j and 3k). As a result of the more homogeneous size and orientation distribution of domains, the modified composites are subject to a more homogeneous distribution of the deformation during loading, preventing high strain localization (Figure SI 12) and leading to the multiple uniformly distributed void nucleation sites in the matrix (Figure 3d and Figure 3g). Finally, we evaluated the effect of the modifications on bridges length distributions. In Mod-(A)-NRCs, the increase in short bridges confirms that the newly generated networks are more constrained and thus rigid, compared to their original counterparts (Figure 3l). Conversely, Mod-(B)-networks have a distribution of bridge lengths that shifts towards larger sizes and becomes multimodal, becoming like those of (A)-networks, suggesting the generation of reinforcing networks with higher local extensibility and hence, higher bridging capability (Figure 3m).

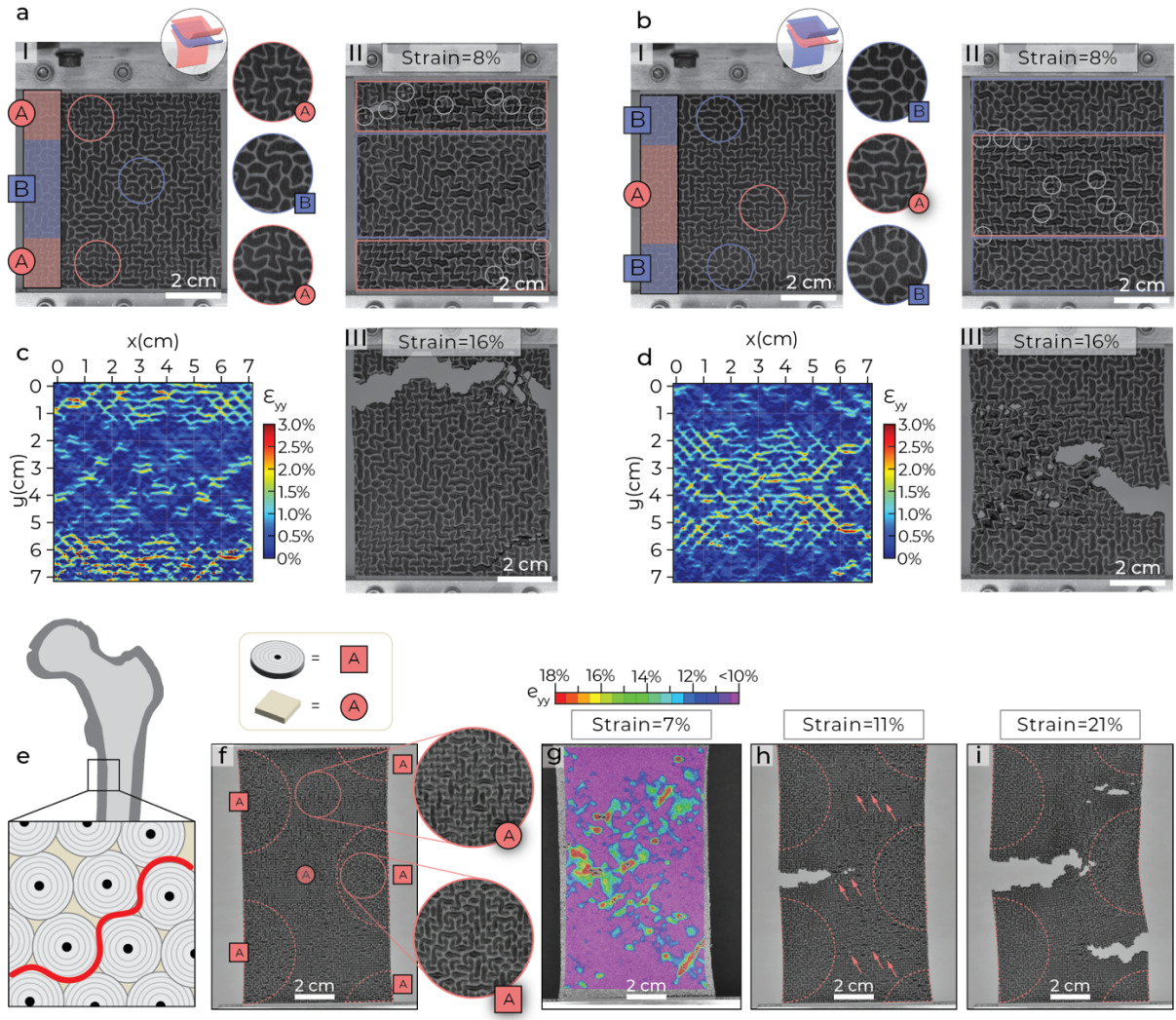

**Figure 4**. **Multi-architecture meso-scale assemblies.** (a,b) Laminate assemblies: (A), Mod-(B), (A) and Mod-(B), (A), Mod-(B) (a and b respectively). The insets highlight differences in reinforcing architecture. Fracture evolution (I, II, III a and b respectively). (c, d) DIC maps at 0.5% strain in laminate assemblies. (e) Sketch of cross section of cortical bone (f) Cortical bone inspired meso-scale assembly. Mod-(A) constitutes osteon-inspired features (dashed red semicircles), (A) constitutes the matrix phase. (g) DIC map at 7% strain and highlighting strain distribution in cortical bone inspired assembly. (h,i) Fracture evolution at 11% and 21% strain (h and i, respectively).

We developed a method to control crack trajectory in network reinforced composites by creating hierarchical microstructures that combine local rules, meso-scale assemblies and macroscale connectivity networks at a constant density. We drew inspiration from biological composites like mother-of-pearl [41–45] and cortical bone [46–48], which deflect incoming cracks and dissipate fracture energy. Our meso-scale assemblies feature rational designs of 'strong and tough' network portions combined with 'soft' network portions. We created two laminate configurations with complementary meso-scale arrangements (Figure 4a I and Figure 4b I, respectively) and found that the (A)-NRCs domains carry most of the strain regardless of their spatial arrangement. For an applied 0.5% strain, (A)-NRCs domains are subject to ~0.8% strain whereas Mod-(B)-NRCs domains experience as little as 0.3% strain (Figure 4c and Figure 4d). We can thus control the fracture trajectory through domain assembly, since fracture nucleates (Figure 4a II and Figure 4b II) and propagates (Figure 4a III and Figure 4b III) in 'soft' domains. These properties are also consistent with crack propagation observed in single edge notch tension tests (SENT) (Figure SI 7). We take inspiration from the cross section of cortical bone, composed of tightly packed osteons, enveloped by the cement lines, specifically designed to arrest and guide incoming cracks on tortuous trajectories (Figure 4e). [49–51] In our cortical bone-inspired assembly we embedded strong and tough osteon-inspired high coordination domains, in a floppy and low coordination matrix domain (Figure 4f). At 7% strain, it is already visible how the strain localizes in the floppy portions of the composite (Figure 4g), leading to fracture nucleation in the central matrix area (left side, Figure 4h), that is then arrested as it approaches the opposite osteon-domain (right side, Figure 4h). Meanwhile, crack nucleation above and below the plane of propagation initiates the desired process of re-nucleation and re-direction of the fracture, critical to deflect its trajectory (red arrows, Figure 4h) and to successfully shield the osteon domains (Figure 4i).

**Conclusions**

In this study, we developed architected composite materials that exhibit a high degree of hierarchical order through material design. By utilizing a virtual growth algorithm, we manipulated the local connectivity between isodensity tiles, resulting in the formation of larger meso-structures, which were merged to create sample-sized assemblies with predetermined spatial arrangements. This approach enabled tailoring the mechanical and fracture properties of the architected composites, at the local and global scales. We envision that the use of different sets of starting tiles and the combination of different reinforcing- and matrix phases, will allow to fine-tune the activation of desired reinforcement and fracture energy dissipation mechanisms. Building on our proof-of-concept observations, we hypothesize that controlling the spatial arrangement and continuity between the soft and hard phases can be used to prevent interfacial failure, while their intentional design can facilitate the precise spatial distribution of fractures in architected composites.

**Acknowledgements**

The authors thank P. Arakelian, K. Liu, T. Zhou, C. McMahan, and J. Boddapati for the fruitful discussions. The authors acknowledge MURI ARO W911NF-21-S-0008 for the financial support. T.M. acknowledges the Swiss National Science Foundation for the financial support.

**Supplementary Information of:**

**Control of Mechanical and Fracture Properties in Two-phase Materials Reinforced by Continuous, Irregular Networks**

Tommaso Magrini*, Chelsea Fox, Adeline Wihardja, Athena Kolli, Chiara Daraio*

*Engineering and Applied Science, California Institute of Technology, Pasadena, CA 91125, USA*

*magrini@caltech.edu,  *daraio@caltech.edu

**Legend**

Experimental Section

Supplementary Figures: Figure SI 1- Figure SI 12

Supplementary Discussion 1- 5

Supplementary References

**Experimental Section**

*Sample Fabrication |* Samples were generated using the virtual growth algorithm described by Liu et. al.28 and further described in the Supplementary Information (Supplementary Discussion 1). The virtual growth algorithm provides a PNG file of the sample architecture, which is then edited using Adobe Illustrator to smoothen all tile connections, ensuring the same volume fraction of reinforcing phase and matrix phase in each sample. Finally, each phase of the sample is extruded and converted into a separate STL file for printing. The specimens are then printed using a Polyjet printer Stratasys Objet500 Connex3, that has a lateral resolution of 40-85 µm.1 The reinforcing phase and matrix phases are printed from Stratasys® VeroWhite Polyjet Resin and Stratasys® TangoBlack Polyjet Resin, respectively.

*Mechanical Characterization |* Uniaxial tension tests were performed on plate geometries of the additively manufactured polymeric composites, with dimensions of 75 mm x 75 mm x 5 mm. An Instron E3000 (Instron, USA) with a 5kN load cell was used to apply a small preload followed by a quasi-static tensile loading at a rate of 2 mm/min. The measured force and displacements were then used to calculate the tensile engineering stress and strain. The experiments were recorded using a Nikon D750 camera (Nikon, USA) with a Nikkor 120 mm f/4 lens (Nikon, USA) at a rate of 1 frame per second.

*Digital Image Correlation |* The same camera setup was used to perform 2D digital image correlation (DIC) on equivalent sets of samples. Samples were painted white using flat white spray paint and then speckled using flat black spray paint such that each speckle was approximately 0.1 to 0.3 mm in diameter and would take up approximately 3x3 to 5x5 pixels of each image. VIC-2D digital image correlation software (Correlated Solutions, USA) was used to calculate the displacements and the resulting Lagrangian strain fields across the different substructures, using a subset size of 31 and a step size of 2, which captured the large global deformation while allowing for sufficient resolution of the local deformation.

**Supplementary Figures**

**Figure SI 1**

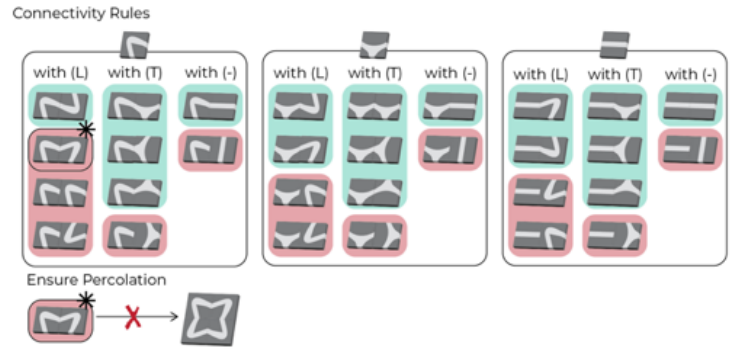

**Figure SI 1:** Connectivity rules for (L), (T) and (-) tiles, with allowed connections (green shade) and prevented connections (red shade).

**Figure SI 2**

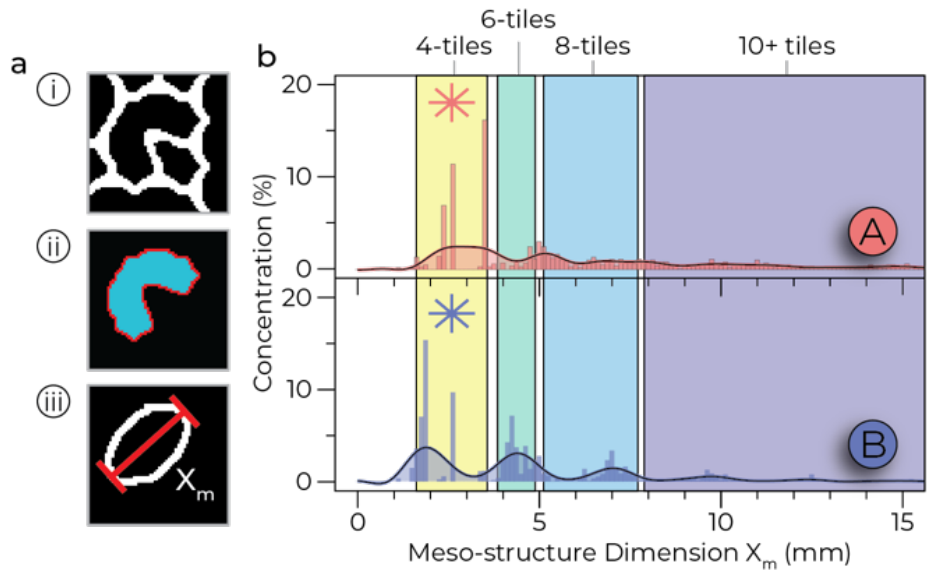

**Figure SI 2:** (a) Meso-structure dimension analysis using elliptic fit to find Xm (meso-structure size), performed by image analysis.[2] (b) Meso-structure size distributions for (A) and (B) architectures relating Xm dimension to concentration percent. Yellow, green, cyan and blue represent 4, 6, 8, and 10+ tiles meso-structures, respectively.

**Figure SI 3**

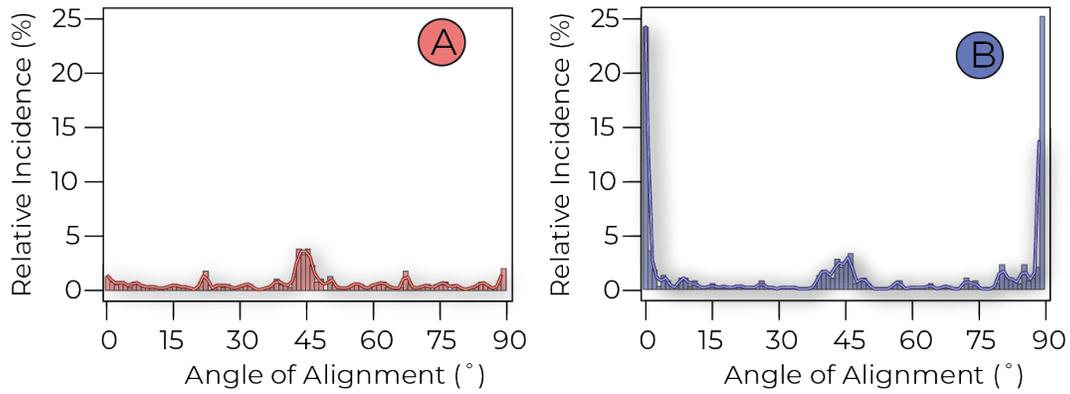

**Figure SI 3:** (a,b) Meso-structure angle distribution in (A) and (B) networks (a and b respectively).

**Figure SI 4**

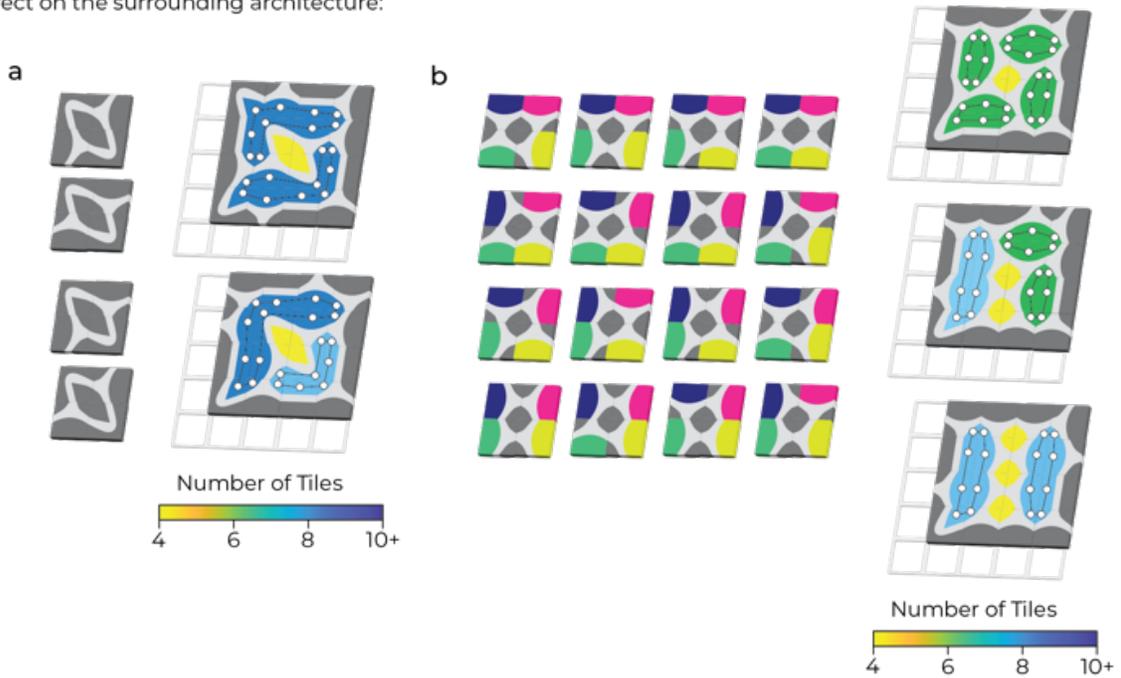

**Figure SI 4:** (a) Meso-structure distributions around (L)-dominated 4-node substructures. (b) Substructure distributions around (T)-dominated architectures.

**Figure SI 5**

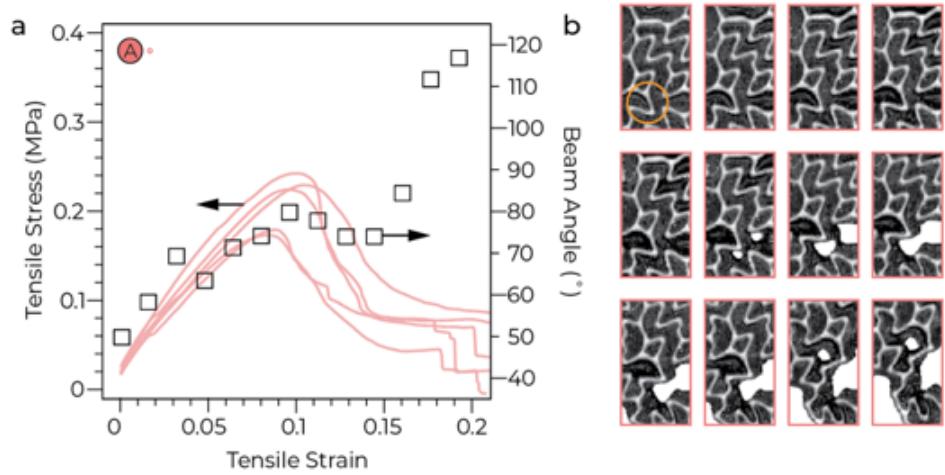

**Figure SI 5:** (a) Engineering tensile strain-stress diagram plotted with opening angle of (L) tile as tensile loading is applied. (b) Progression of (L) tile angle opening under tensile loading.

**Figure SI 6**

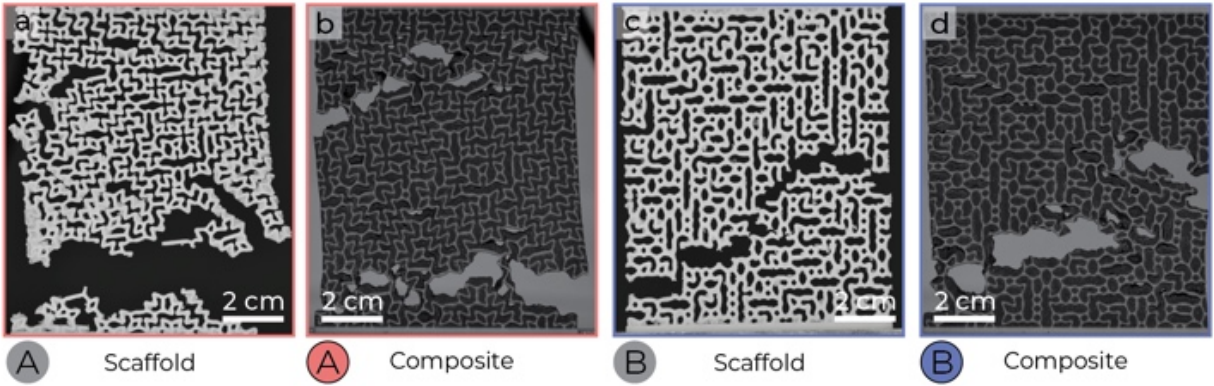

**Figure SI 6:** Comparison of reinforcing network and composite failure locations under tensile loading. (a) (A) reinforcing network and (b) (A)-NRC showing the same failure locations. (c) (B) reinforcing network and (d) (B)-NRC showing the same failure locations.

**Figure SI 7**

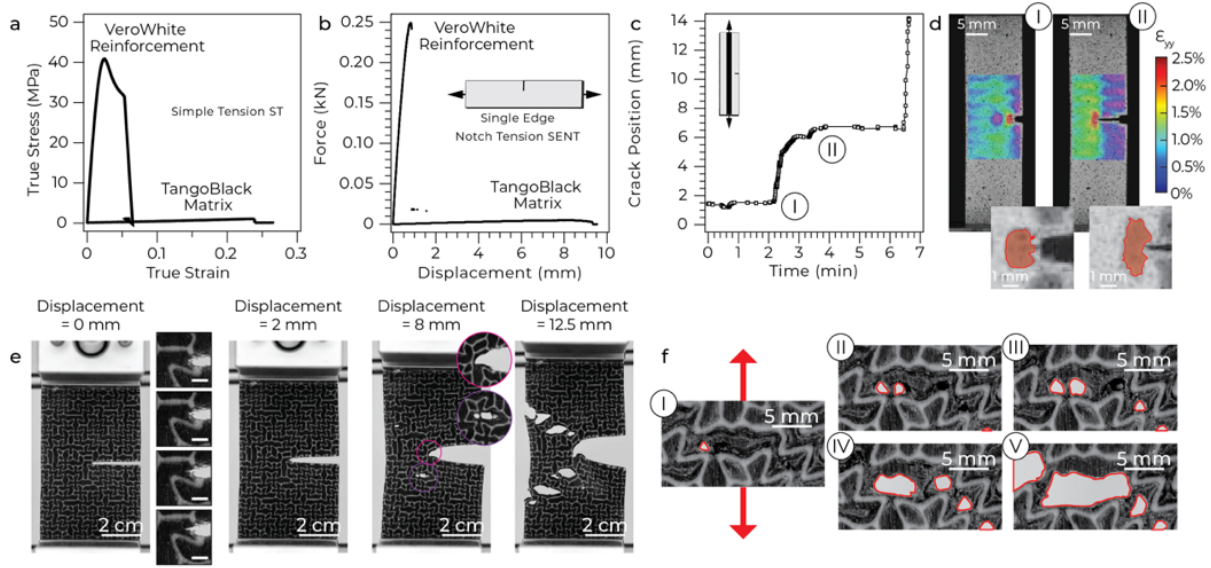

**Figure SI 7:** (a) Simple tension experiment performed on VeroWhite and TangoBlack specimens. (b) Single edge notch tension (SENT) experiment performed on VeroWhite and TangoBlack specimens at a loading rate of 1mm/min. (c) Evolution of the crack position over time during the fracture of a VeroWhite-TangoBlack-VeroWhite specimen, confirming sequential events of crack arrest at the interface (I and II). The crack position has been retrieved by image analysis.[2] (d) Digital Image Correlation (DIC) maps of the $e_{yy}$ strain during the test reported in (c). The inset confirms the sequential formation of a plastic zone in the VeroWhite portions subject to local yielding, as also observed in literature.[2,3] (e) Optical photographs during a controlled fracture experiment (SENT geometry) of a network reinforced composite loaded at 1mm/min (initial crack length/specimen width ~ 0.5). Sequential details display the instantaneous propagation of a crack at small displacements internally to the first mesostructure, within the TangoBlack matrix (scale bar = 1 mm). Circular insets highlight the event of fracture arrest at the first soft to hard interface and the nucleation of a void in a nearby mesostructure. (f) Optical photographs highlighting the sequence of craze formation, fibril elongation and fracture within the TangoBlack matrix phase, internally to a mesostructure.

**Figure SI 8**

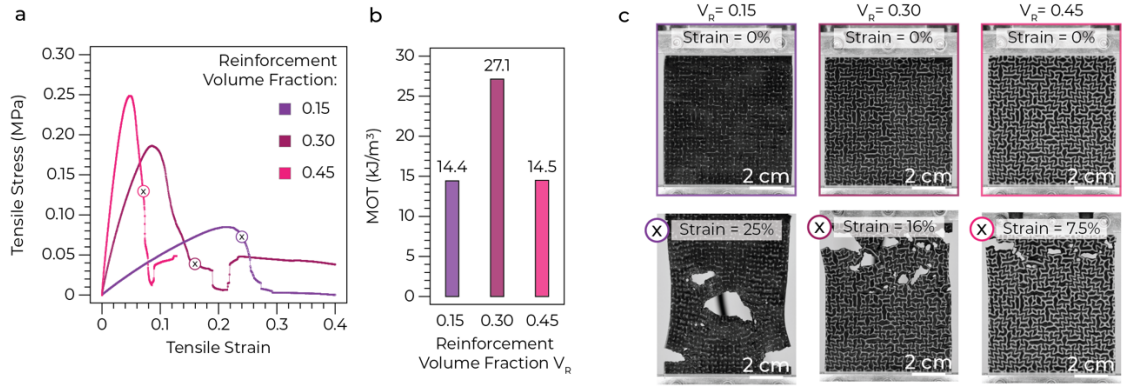

**Figure SI 8:** (a) Stress strain curves recorded on A-NRCs with a different reinforcement volume fraction $V_R$, ranging from 0.15 to 0.45. (b) Modulus of Toughness (MOT) as a function of the reinforcement volume fraction. (c) Optical photographs depicting the evolution of the fracture in specimens with increasing reinforcement volume fraction 0.15 (left), 0.30 (center) and 0.45 (right).

**Figure SI 9**

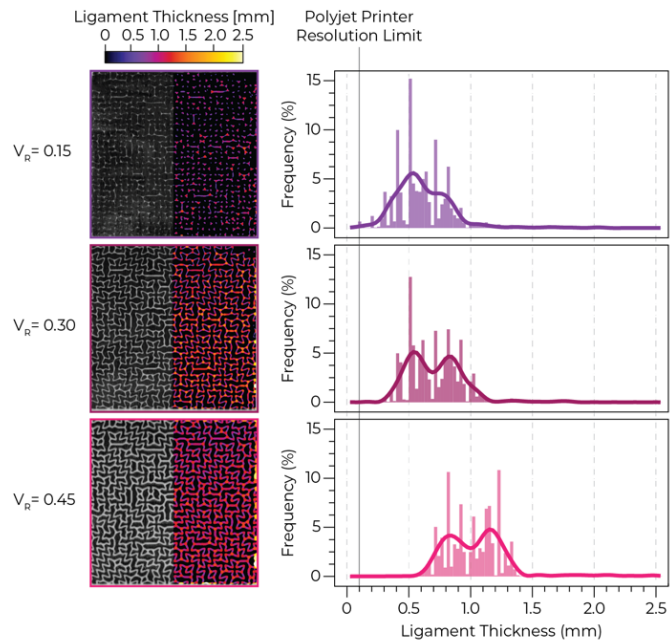

**Figure SI 9:** (Left) Optical photographs of composites with increasing reinforcement volume fraction $V_R$ ranging from 0.15 to 0.45. The ligament thickness map is obtained with the plugin

'local thickness' [3] in the open-source image analysis software Fiji[2] and it is then overlayed on the images. (Right) Ligament thickness distribution for each specimen (bars), smoothing of the distributions measured for each specimen (solid lines).

**Figure SI 10**

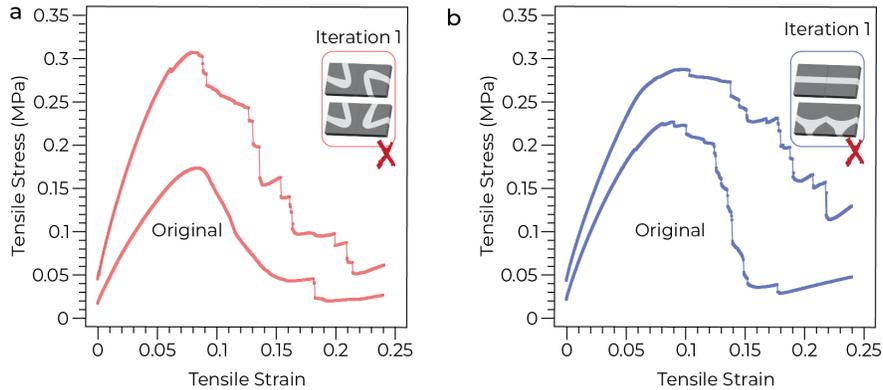

**Figure SI 10:** Engineering tensile stress-strain diagrams showing connectivity rule modification effects. (a) Original (A) architecture (lower curve) comparison with first iteration (upper curve) of connectivity rule modifications as shown in upper right corner. (b) Original (B) architecture (lower curve) comparison with first iteration (upper curve) of connectivity rule modifications as shown in upper right corner.

**Figure SI 11**

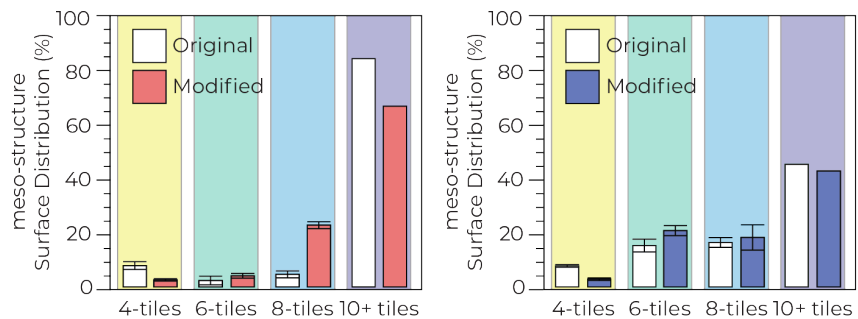

**Figure SI 11:** Variations in distribution of meso-structure sizes for (A)-NRCs and Mod-(A)-NRCs (left) and (B)-NRCs and Mod-(B)-NRCs (right), measured by image analysis.[2] Yellow, green, cyan and blue represent 4, 6, 8, and 10+ tiles meso-structures, respectively, before (white bars) and after (red and blue bars) connectivity rule modifications.

**Figure SI 12**

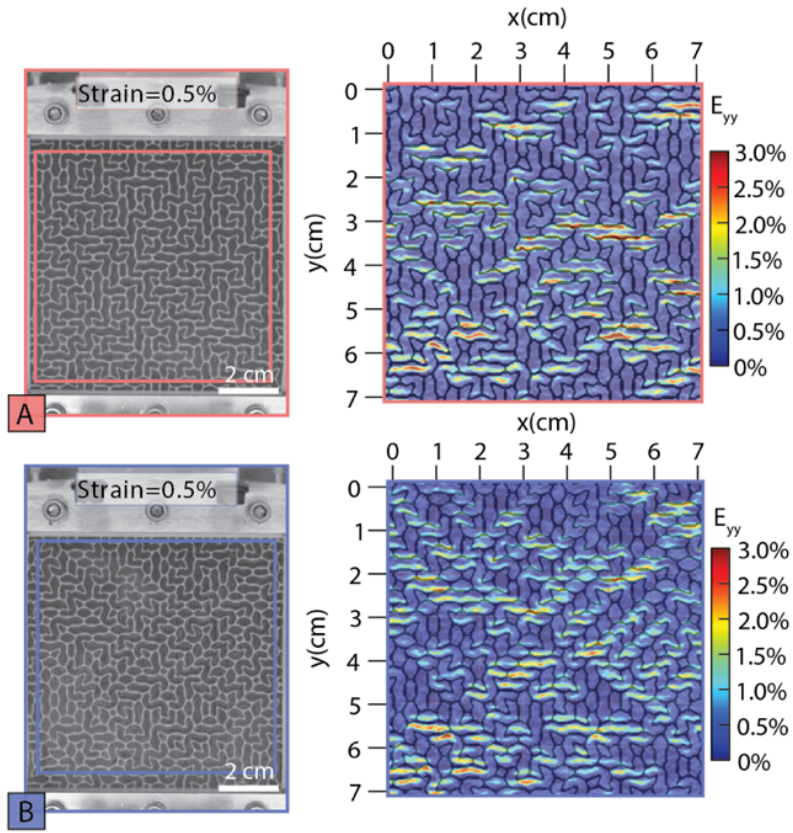

**Figure SI 12:** Images (left) and digital image correlation maps (right) recorded at 0.5% strain of (A)* and (B)* composites (top and bottom, respectively).

**Supplementary Discussion 1 | The growth algorithm**

The program used to design the composites iteratively grows irregular metamaterials over a predefined area by selecting spatial sites, defined by two coordinates, and assigning tiles to these coordinates [Liu, Sun, Daraio, Science, 377(6609), 975-981 (2022)].[4] The entropy of each site, defined as the number of available connections that can still be formed, a number that ranges from 1 to 4, is evaluated, and the sites with the lowest entropy are filled first. The algorithm then randomly selects tiles based on their initial concentration or availability, reported often as 'frequency hints' (Figure 1b, left). This parameter is provided as an input by the user. To ensure a smooth growth without local defects or discontinuity between the reinforcing and matrix phases, the connectivity of each tile is governed by a set of connectivity rules, also known as adjacency rules (Figure SI 1). As the structure grows and the sites are occupied, each remaining available site can only be occupied by a limited number of tiles, arranged in a limited number of rotational configurations (Figure 1b, center). The growth algorithm then proceeds with the random selection of one of the allowed tiles until it ensures the complete filling of the predefined area (Figure 1b, right). For this study, a specific (L) to (L) connection is not allowed, to prevent the formation of secondary disconnected architectures (Figure SI 1).

**Supplementary Discussion 2 | The Reinforcing Networks**

Networks are mathematical models that describe systems of nodes connected by edges. The rules that govern how nodes connect ultimately determine the network's characteristics, distinguishing an irregular network, which follows a set of rules to form an irregular pattern, from a random network, where nodes are connected randomly. Relying on the theoretical frameworks developed for continuous random networks, that describe the relationship between networks architecture and mechanical properties, we can characterize the reinforcing networks generated using the virtual growth algorithm based on their average coordination <R>, the mean number of connections each node has, and their bridge length, defined as the distance between two nodes. In this context, among the different models developed, covalent networks have been extensively studied to describe amorphous solids and glasses, composed by atoms with different coordination, analogous to our system tiles. The average coordination of covalent networks <R> can be measured through constraint counting, estimating the network's stability and identifying floppy modes, independent deformations that occur with no cost in energy. The critical coordination <$R_C$> marks the transition between rigid and floppy behavior and can be used to predict the mechanical behavior of the network.[5,6]

**Supplementary Discussion 3 | The Mechanical Behavior of the Reinforcing Networks**

To confirm this description of the failure mechanisms in (L)-dominated and (T)-dominated architectures, we performed control experiments on the reinforcing networks and compared the results with the behaviour of the composites.

(L)-dominated architectures: the experiments on the reinforcing networks consolidated our description of a transition from a bending- to a stretching-dominated behaviour. As the strain exceeded 20% (figure 3a, grey solid line and grey arrows, and Figure SI 6) we observed the sequential fracture of the reinforcing network ligaments. Finally, these experiments allowed us to shed light on the role of the matrix during fracture. The reinforcing networks display a significantly lower strength than the composites: this suggests that the matrix has a key role during loading in resisting against the deformation of the meso-structures, requiring higher forces and thus increasing the total amount of energy that is needed to cause composite failure (Figure 3a, black solid line composite, grey solid line reinforcing network). Nonetheless, the matrix has a marginal role in influencing the trajectory of the cracks, that can almost perfectly overlap between the composite and the reinforcing network.

(T)-dominated architectures: To consolidate our description of their mechanical behavior, we performed control experiments on the reinforcing network of (B) architectures. As expected, they display a primarily stretching-dominated behavior, that features sequential failure events at lower strain values (Figure 3b, grey arrows, Figure SI 6). As observed in the (L)-dominated architectures, the crack trajectory can almost perfectly overlap between the composite and the reinforcing network.

**Supplementary Discussion 4 | Iterative modification design process**

With our approach, we display that without changing the shape or the volume fraction of the constitutive tiles, and without any shape optimization, it is possible, simply by changing the rules that govern the connectivity of tiles, to tune the mechanical properties of these architected materials. Although this study focused on improving the energy dissipated during fracture of the architectures, the iterative process of modifying connectivity rules and characterizing the mechanical response allowed us to identify which set of rules can be modified to increase or to decrease various other mechanical properties. The two specific mechanical properties we examined during the iterative process were strength and strain-to-failure, which combined to give an improved fracture energy dissipation.

We began by removing the most prominent defect in each architecture. The tile distributions are directly responsible for the frequency of the type of defect in each architecture, and the most prominent defect is the one that occurs most often in an architecture. For (A), that was certain (L) to (L) connections, while for (B), certain (T) to (T) connections were the most problematic. Although the modified architectures (A) and (B) were generated using the exact same frequency hints as their original counterparts, the modification of the connectivity rules, and thus the possible substructures that could be formed, influenced their final composition: the modified designs (A_NO LL) and (B_NO TT) have a composition of 50% (T), 19% (L), and 31% (-) and 68% (T), 22% (L), and 10% (-), respectively.

We then performed the same mechanical testing and characterization of the modified (A) and (B) composites as described for the original composites. For both composites, the removal of the most prominent defects resulted in an increase in the modulus of toughness, but the mechanism behind the increase was different for (A) and (B) (Figure SI 7). The increase in modulus of toughness for (A) was due to an increase in strength, while the increase in modulus of toughness for (B) was due to an increase in strain-to-failure. From this, we concluded that removal of (L) to (L) connections is responsible for strength and that removal of (T) to (T) and (-) to (-) connections is responsible for strain-to-failure.

To test this hypothesis, we then removed the (-) to (-) connections from (A) to improve its strain-to-failure (resulting in 37% (T), 58% (L), and 5% (-)), and the (L) to (L) connections from (B) (resulting in 83% (T), 9% (L), and 8% (-)) to improve its strength. We again performed the same mechanical testing and characterization of the newly modified (A) and (B) architectures, and as expected, observed that the removal of (-) to (-) connections improved the strain-to-failure from the original (A) architecture, while removal of (L) to (L) connections improved the strength from the original (B). However, unlike the first modification, the second modifications did not have as great of an impact on the increase in modulus of toughness.

The improvements in strength and strain-to-failure are due to the tile distributions and thus meso-structure populations that result from the connectivity rules applied. Since energy dissipation during fracture is optimized by a combination of strength and strain-to-failure, we then decided to combine all the modulus of toughness improvements into a final set of connectivity rules that were applied to the original (A) and (B) architectures.

**Supplementary Discussion 5 | The Polydispersity Index (PDI)**

Borrowing the concept from the field of polymer science, a measure of the polydispersity of a polymer is the polydispersity index (PDI), defined as the ratio of Mw (weighted average mass) over Mn (number average mass). The parallel between polymers and irregular architected materials is apparent in this case; each tile can be seen as one monomer, and each formed substructure can be seen as one polymer chain. The total amount of tiles (or *monomers*) is fixed by the total extension of the material and the final assembly will therefore be a collection of differently sized meso-structures (or *polymer chains*). In this context, we propose here that comparing the PDI of each architecture can be a quick method to evaluate the homogeneity of each architecture.

The PDI is calculated following Equation 1:

$$PDI = M_W/M_N \quad (1)$$

In which $M_W$ and $M_N$ are given by Equation 2 and 3 respectively:

$$M_W = \sum(Area * Counts^2) / \sum(Area * Counts) \quad (2)$$

$$M_N = \sum(Area * Counts) / \sum(Counts) \quad (3)$$

$Area$ is the enclosed area of a n$^{th}$ meso-structure and $Counts$ is the frequency (the number of times) it gets measured.

As it is defined, the PDI is larger than 1.0 and the closer it is to 1.0, the more homogeneous the architecture is. As a result of more stringent connectivity rules, that bias the growth of architectures capable of higher energy dissipation during fracture, we observed across the design space an overall decrease of the PDI, suggesting that the modified architectures become more homogeneous (Table 1).

Table 1:

| Architecture # | (T) | (-) | (L) | PDI Original | (T)* | (-)* | (L)* | PDI Modified* |
| --- | --- | --- | --- | --- | --- | --- | --- | --- |
| 1 | 62 | 18 | 20 | 1.78 | 63 | 18 | 19 | 1.41 |
| 2 | 60 | 29 | 11 | 1.81 | 65 | 18 | 17 | 1.37 |
| 3 | 57 | 6 | 37 | 2.08 | 64 | 12 | 24 | 1.58 |
| 4 | 42 | 45 | 13 | 1.85 | 58 | 22 | 20 | 1.39 |
| 5 | 19 | 2 | 79 | 1.80 | 49 | 26 | 25 | 1.36 |
| 6 (A) | 35 | 10 | 55 | 2.09 | 54 | 19 | 27 | 1.41 |
| 7 | 36 | 28 | 36 | 2.27 | 53 | 24 | 21 | 1.51 |
| 8 (B) | 80 | 10 | 10 | 1.81 | 71 | 11 | 18 | 1.45 |

**Supplementary References:**